\documentclass[preprint,showpacs,aps,superscriptaddress,nofootinbib]{revtex4}
\usepackage{graphicx}
\begin{document}

\title{Quantization of Friedmann cosmological models with two fluids:
dust plus radiation}

\author{N. Pinto-Neto }\email{nelsonpn@cbpf.br}
\affiliation{Centro Brasileiro de Pesquisas F\'{\i}sicas, \\
Coordena\c c\~ao de Cosmologia, Relatividade e Astrof\'{\i}sica: ICRA-BR,\\
Rua Dr.\ Xavier Sigaud 150, Urca 22290-180 -- Rio de Janeiro, RJ -- Brasil}

\author{E. Sergio Santini}\email{santini@cbpf.br}
\affiliation{Centro Brasileiro de Pesquisas F\'{\i}sicas, \\
Coordena\c c\~ao de Cosmologia, Relatividade e Astrof\'{\i}sica: ICRA-BR,\\
Rua Dr.\ Xavier Sigaud 150, Urca 22290-180 -- Rio de Janeiro, RJ -- Brasil}
\affiliation{ Comiss\~ao Nacional de Energia Nuclear \\
Rua General Severiano 90, Botafogo 22290-901 -- Rio de Janeiro, RJ -- Brasil}

\author{F. T. Falciano}\email{ftovar@cbpf.br}
\affiliation{Centro Brasileiro de Pesquisas F\'{\i}sicas, \\
Coordena\c c\~ao de Cosmologia, Relatividade e Astrof\'{\i}sica: ICRA-BR,\\
Rua Dr.\ Xavier Sigaud 150, Urca 22290-180 -- Rio de Janeiro, RJ -- Brasil}

\date{\today}

\begin{abstract}

The causal interpretation of quantum mechanics is applied to a homogeneous
and isotropic quantum universe, whose matter content is composed by
non interacting dust and radiation. For wave functions which are eigenstates
of the total
dust mass operator, we find some bouncing quantum universes
which reachs the classical limit for scale factors much larger than
its minimum size. However these wave functions do not have unitary evolution.
For wave functions which are not eigenstates of the dust total mass
operator but do have unitary evolution,
we show that, for flat spatial sections, matter can be created as a quantum
effect in such a way
that the universe can undergo a transition from an exotic matter dominated
era to a matter dominated one.

\end{abstract}
\maketitle

\section{Introduction}

The Bohm-de Broglie (BdB) interpretation \cite{bohm1}\cite{bohm2}\cite{hol}
has been sucessfully applied to quantum minisuperspace models
\cite{vink,bola1,kow,hor,bola2,fab,fab2}, and
to full superspace \cite{must} \cite{cons} \cite{tese}. In the first case, it
was discussed the classical limit, the singularity problem, the
cosmological constant problem, and the  time issue.  It was shown in
scalar field and radiation models for the matter content of the early
universe that quantum effects driven by the quantum potential can
avoid the formation of a singularity through a repulsive quantum
force that counteract the gravitational attraction. The
quantum universe usually reach the  classical limit for large scale
factors.  However, it is possible to have small classical universes and
large quantum ones:  it depends on the  state vector and on initial
conditions \cite{fab}. It was also shown that  the quantum evolution of
homogeneous hypersurfaces form the same four-geometry independently on
the choice of the lapse function \cite{bola1}.

In the present work we study the minisuperspace model given by a
quantum Friedmann-Lema\^{\i}tre-Robertson-Walker (FLRW) universe filled with dust
and radiation
decoupled  from each other. We write down the hamiltonian that comes from the
velocity potential Schutz formalism \cite{schutz1}. After
implementing a canonical transformation,
the momentum associated to the radiation fluid
$p_{T}$ and to the dust fluid $p_{\varphi}$ appear linearly in the superhamiltonian
constraint. Both can be associated to time parameters, but physical reasons
and mathematical simplicity led us to choose
the coordinate $T$ associated with $p_{T}$ as the time parameter. This is equivalent
to choose the (reversed) conformal time. We quantize this system
obtaining a Schr\"odinger-like equation. We analyze its time dependent
solutions applying the BdB interpretation in order to study the
scale factor quantum dynamics.

We first consider an initial quantum state given by a gaussian superposition of the
scale factor which is an eigenstate of the total dust mass operator
(matter is not being created
nor destroyed in such states), and we compute the solution at a general subsequent
time by means of
the propagator approach. We calculate the bohmian trajectories for the scale factor.
For flat and negative curvature spatial sections, we find that the quantum solutions for
the scale factor reach the classical behaviour for long times, but do not present
any initial singularities due to quantum effects.
In the same way, in the case of positive curvature spatial sections, the classical
initial and final singularities are removed due to quantum effects, and
the scale factor oscillates between a minimum and a maximum size. For large
scale factor,
the classical behaviour is recovered. However, such eigenfunctions of the
total dust mass operator do not have unitary evolution. This led us to
consider an initial state given by gaussian superpositions of
the total dust matter content. In this situation, dust and radiation can be
created and destroyed.
We calculate general solutions for flat, negative and positive curvature
spatial sections.
In particular, for flat spatial sections, we construct a wave packet
whose quantum trajectories represent universes which
begin classically in an epoch where the dust matter has negative energy density
(exotic dust matter), evolving unitarily to
a configuration where
quantum effects avoid the subsequent classical big crunch singularity,
performing a graceful exit to an expanding classical model filled with
conventional matter and radiation. There is thus a transition from an exotic
matter era to a conventional matter one due to quantum effects.

This paper is organized as follows. In section \ref{bdbs} we synthesize the basic
features of the Bohm-de Broglie interpretation  of quantum mechanics, which will be
necessary to interpret our quantum model studied in other sections. In section
\ref{drs}, we briefly summarize the velocity potential Schutz formalism, and
we apply it
to construct the hamiltonian of the FLRW universe filled with two perfect fluids,
which are dust and radiation. We then review and analyze the
classical features of the two perfect fluids FLRW model in order to have the
results to be contrasted with the quantum models of the following sections.
In section \ref{1f}, we present some new results concerning
the existence of singularities in the quantization of the one fluid case.
We show that, when the fluid is radiation, all quantum solutions do
not present singularities.
In section \ref{quantum}, we quantize the model with two fluids, and we
compute the solutions of the
Schr\"odinger like equation for two different initial conditions: the first
being an eigenstate of the total dust matter operator, and the second a gaussian
superposition of
total dust matter eigenstates. We interpret the solutions according to the BdB
view and we
develop the main results of the paper. Section \ref{conclu} is for discussion
and conclusions.

\section{The Bohm-de Broglie interpretation of quantum mechanics}\label{bdbs}
In this section, we briefly review the basic principles of the
Bohm-de Broglie (BdB)
interpretation of quantum mechanics. According to this causal interpretation,
an individual physical system comprises a wave  $\Psi(x,t)$, which is a solution
of the Schr\"odinger equation, and a point particle that follows a trajectory
${x}(t)$, independent of observations, which is solution of the Bohm
guidance equation

\begin{equation}\label{bohmg}
p=m\dot{x}=\nabla S(x,t)|_{x=x(t)} ,
\end{equation}
where $S(x,t)$ is the phase of $\Psi$. In order to solve Eq.(\ref{bohmg}),
we have to specify the initial condition $x(0)=x_0$. The uncertainty in the
initial conditions define an ensemble of possible motions,
\cite{bohm1}\cite{bohm2}\cite{hol}.

It is sufficient for our purposes to analyze the Schr\"odinger equation for a
non relativistic particle in a potential $V(x)$, which, in coordinate representation, is

\begin{equation}\label{s}
 i\frac{\partial\Psi(x,t)}{\partial t}=
 \biggl[-\frac{1}{2m}\nabla^2 +V(x)\biggr]\Psi(x,t) .
\end{equation}
Substituting in (\ref{s}) the wave function in polar form, $\Psi=A \exp (iS)$, and
separating into real and imaginary parts, we obtain the following two equations for the
fields $A$ and $S$

\begin{equation}\label{equacaoHJ}
\frac{\partial {S}}{\partial t}+\frac{\left(\nabla S\right)^2}{2m} +
V-\frac{1}{2m}\frac{\nabla^2 A}{A}=0 ,
\end{equation}

\begin{equation}
\frac{\partial A^2}{\partial t}+\nabla.\left(A^2\frac{\nabla S}{m}\right)=0 .
\end{equation}
Equation (\ref{equacaoHJ}) can be interpreted as a Hamilton-Jacobi type equation for a
particle submitted to a potential, which is given by the classical potential $V(x)$
plus a {\it quantum potential} defined as
\begin{equation}\label{qpote}
Q\equiv -\frac{1}{2m}\frac{\nabla^2 A}{A} .
\end{equation}
It is possible to verify that the particle trajectory $x(t)$ satisfies the equation of
motion

\begin{equation}
m\frac{d^2 x}{dt^2}=-\nabla V - \nabla Q   .
\end{equation}

The classical limit is obtained when $Q=0$. The
BdB interpretation does not need a classical domain outside the quantized system
to generate the physical facts out of potentialities.
In a real measurement, we do not see superpositions of the pointer apparatus because
the measurement interaction causes the wave function to split up into a set  of
non overlapping packets. The particle will enter in one of them, the rest being empty,
and it will be influenced by the unique quantum potential coming from the sole non zero
wave function of this region. The particle cannot pass to another branch of the
superposition because they are separated by regions where $\Psi=0$,
nodal regions.

In section \ref{quantum}, the FLRW minisuperspace model containing dust and
radiation as two perfect decoupled fluids will be quantized. A preferred time
variable can be chosen
as one of the velocity potencials associated to the fluids (radiation),
yielding a Schr\"odinger like equation.
Then the BdB interpretation of our
quantum model runs like in Ref. \cite{bola2}, in close analogy to the
non relativistic particle model described above.
In the present case, however, the scale factor of the universe will not be
the only degree of freedom: the velocity potential associated with the dust field
and its canonical momentum, interpreted as the dust total mass, are also present.
They satisfy a Hamilton-Jacobi equation modified by an
extra term, the quantum potential, so that their time evolution will be
different from the classical one. The main features of this classical
model we describe in the following section.

\section{Classical dust plus radiation model in the velocity potential Schutz formalism}
\label{drs}
We start by considering a perfect fluid in a FLRW universe model.
The line element is given by

\begin{equation}\label{metric}
ds^{2}=-N^{2}dt^{2}+a^{2}\left(t\right)\gamma_{ij}dx^{i}dx^{j}
\end{equation}
where $N$ is the lapse function, $a$ is the scale factor, and $\gamma_{ij}$ is
the metric of  the three-dimensional  homogeneous isotropic static spatial section
of constant curvature $\kappa=1, 0,$ or $-1$.

Following the Schutz's canonical formalism to describe the relativistic dynamics of
a perfect fluid in interaction with the gravitational field \cite{schutz1}, we
introduce the five velocity potentials, $ \alpha, \beta, \theta, \varphi$ and  $s$.
The potentials $\alpha$ and $\beta$, which describe vortex motion, vanish in the FLRW
model because of its symmetry.
The potential $s$ is the specific entropy and $\theta$ can be related with the
temperature of the fluid. By now $\varphi$ works only as a mathematical tool.

The four-velocity of the fluid is obtained from the velocity potentials as

\begin{equation}
\mathit{U}_{\nu}=\frac{1}{\mu}\left(\varphi,_{\nu} +\theta\,s,_{\nu}\right),
\end{equation}
where $\mu$ stands for the specific enthalpy.
The four velocity is normalized as
\begin{equation}
g_{\alpha\,\beta}\mathit{U}^{\alpha}\mathit{U}^{\beta}=-1 .
\end{equation}
Using this equation, it is possible to write
the specific enthalpy $\mu$ as a function of the velocity potentials.

The action for a relativistic perfect fluid and the gravitational field
in the natural units $c=\hbar=1$ is given by

\begin{equation}\label{A}
I = -\frac{1}{6l_p^2}\int_{M}d^{4}x\sqrt{-g}\, ^{4}{\cal R}+
\int_{M}d^{4}x\sqrt{-g}\, p+
\frac{1}{3l_p^2}\int_{\partial M} d^{3}x\sqrt{h}h_{ij}K^{ij},
\end{equation}
where $l_p\equiv(8\pi G/3)^{-1/2}$, $G$ is the Newton's constant
(hence $l_p$ is the Planck length in the natural units),
$^{4}{\cal R}$ is the scalar curvature of the
spacetime, $p$ is the pressure of
the fluid, $h_{ij}$ is the three metric on the boundary $\partial M$ of the
4-dimensional
manifold $M$, and $K^{ij}$ its extrinsic curvature.
The velocity potentials are supposed to be functions of $t$ only, in accordance
with the
homogeneity of spacetime. The perfect fluid follows the equation of state
$p=\lambda \rho$.

Substituting the metric (\ref{metric}) into the action (\ref{A}),
using the formalism of Schutz \cite{schutz1} to write the pressure
of the fluid as
\begin{equation}
p= p_{0r}\left[ \frac{\dot{\varphi}+\theta\dot{s}}
{N(\lambda+1)}
\right]^{\frac{\lambda+1}{\lambda}}
\exp{\left(-\frac{s}{s_{0r}\lambda} \right)},
\end{equation}
with $p_{0r}$ and $s_{0r}$ constants, computing the canonical momenta
$p_{\varphi},p_{s}, p_{\theta}$ for the fluid and $p_a$ for the gravitational
field, using the two constraints equations
$p_{\theta}=0, \,\,\, \theta p_{\varphi}=p_s $,
and performing the canonical transformation
\begin{equation}
T=-\frac{p_s}{6^{1-3\lambda}}\exp \left( -\frac{s}{s_{0r}}\right)
p_\varphi^{-(\lambda+1)}\rho_{0r}^{\lambda}s_{0r},
\label{can1}
\end{equation}
and
\begin{equation}
\varphi_N=\varphi + (\lambda + 1)s_{0r} \frac{p_s}{p_\varphi},
\label{can3}
\end{equation}
leading to the momenta
\begin{equation}p_{_T}=6^{1-3\lambda}
\frac{p_\varphi^{(\lambda+1)}}{\rho_{0r}^{\lambda}}
\exp\left(\frac{s}{s_{0r}}\right),
\label{can2}
\end{equation}
and
\begin{equation}
p_{\varphi_N}=p_{\varphi},
\label{can4}
\end{equation}
we obtain for the final Hamiltonian
(see Ref. \cite{Lapshinskii} for details),
\begin{equation}\label{superh}
H\equiv N{\cal H}=N\biggl(-\frac{p_{a}^2}{24a}-6\kappa a+
\frac{p_T}{a^{3\lambda}}\biggr),
\end{equation}
where $N$ plays the role of a Lagrange multiplier whose variation yields
the constraint equation
\begin{equation}\label{constr}
{\cal H}\approx 0,
\end{equation}
where $\approx$ means `weakly zero' (this phase space function is constrained
to be zero, but its Poisson bracket to other quantities is not).
We have redefined $\tilde{a}=\sqrt{V/(16\pi l_p^2)}\; a$ in order for $\tilde{a}$
be dimensioless, and $\tilde{N}=\sqrt{6}N$, where $V$ is the total comoving
volume of the spatial sections. The tilda have been omitted.
Considering now two decoupled fluids, one being radiation ($\lambda_r=1/3$),
and the other dust matter ($\lambda_d=0$), the Hamiltonian reads:
\begin{equation}\label{hrm}
H \equiv N{\cal H}=N\biggl(-\frac{p_{a}^2}{24a}-6\kappa a+
\frac{p_{T}}{a}+p_{\varphi}\biggr)
\end{equation}

The  classical Hamilton equations are:

\begin{equation}
\label{aponto}
\dot{a}=\left\{a,H\right\}=-\frac{N}{12a}p_{a}
\Rightarrow p_{a}=-\frac{12a \dot{a}}{N} ,
\end{equation}

\begin{equation}
\label{13}
\dot{p_{a}}=\left\{p_{a},H\right\}=
N\biggl(-\frac{p_{a}^2}{24 a^2}+6\kappa+
\frac{p_{T}}{a^2}\biggr),
\label{pr14}
\end{equation}

\begin{equation}
\label{conf}
\dot{T}=\frac{N}{a}  ,
\end{equation}

\begin{equation}
\label{cosm}
\dot{\varphi}=\left\{\varphi,H\right\}=N ,
\end{equation}

\begin{equation}\label{15}
\dot{p}_{T}=\dot{p}_{\varphi}=0 \Rightarrow \mbox{$p_{T}$, $p_{\varphi}$ are constants}.
\end{equation}

The superhamiltonian is constrained to vanish due to variation of the Hamiltonian
with respect to the lapse function $N$, ${\cal H}\approx0$,

\begin{equation}\label{ham0}
-\frac{p_{a}^2}{24a}-6\kappa a + \frac{p_{T}}{a} + p_{\varphi} = 0 .
\end{equation}
The constraint (\ref{ham0}) combined with Eqs. (\ref{aponto}) and (\ref{15})
yield the Friedmann's equation

\begin{equation}\label{friedmann}
\left(\frac{\dot{a}}{a}\right)^{2}=N^{2}\left[-\frac{\kappa}{a^{2}}+\frac{1}{6}
\left(\frac{p_{T}}{a^{4}}+
\frac{p_{\varphi}}{a^{3}}\right)\right]
\end{equation}
Note that the conjugate momenta $p_T$ and $p_{\varphi}$, classical constants of motion,
can be identified to the total content of dust and radiation in the universe:

\begin{equation}
p_{\varphi}=16\pi Ga^{3} \rho_{m} ,
\end{equation}
\begin{equation}
p_{T}=16\pi Ga^{4} \rho_{r}.
\end{equation}

Note also that Eq.(\ref{cosm}) implies that $d\varphi=Ndt$, hence $\varphi$ is
cosmic time, while Eq.(\ref{conf}) yields $adT=Ndt$ so $T$ is conformal time.
Consequently, choosing $N=1$ means taking coordinate time $t$ as cosmic time $\varphi$,
while choosing $N=a$ imposes coordinate time to be conformal time $T$.
Explicit analytic solutions of Eqs.(\ref{aponto},\ref{pr14},\ref{15},\ref{friedmann})
can be obtained only in the gauge $N=a$. In this gauge, besides the constraint
(\ref{friedmann}) with $N=a$, we obtain the simple second order equation,
\begin{equation}
\label{2order}
a''+\kappa a=\frac{p_{\varphi}}{12},
\end{equation}
where a prime means differentiation with respect to conformal time, which we denote $\eta$ from now on.
The solutions read:

\begin{equation}\label{cdr}
a = \left\{ \begin{array}{ll}
 \left(\frac{2a_{eq}}{\eta_{eq}^{2}}\right)\left[1-\cos( \eta )+
 \eta_{eq}\sin( \eta)\right]  & \;\; ; \kappa=1 ,\\
\\
a_{eq}\left[2\frac{\eta }{\eta_{eq}}+\left(\frac{\eta }{\eta_{eq}}\right)^{2}\right]
& \;\; ; \kappa=0 ,\\
\\
\left(\frac{2a_{eq}}{\eta_{eq}^{2}}\right)\left[\cosh( \eta )+
\eta_{eq}\sinh( \eta)-1\right] & \;\; ; \kappa=-1 .
\end{array} \right.
\end{equation}
The quantity $a_{eq}$ is defined to be the value of the scale
factor at the equilibrium time where $\rho_{m}=\rho_{r}$, and
$\eta_{eq}^{2}=3/\left(2\pi\, G \, \rho_{r}a^4\right)=24\,
a_{eq}/\mid  p_{\varphi}\mid $.

As we will see in section (\ref{quantum}), the presence of quantum effects can create
exotic dust matter content. Hence, for comparison, we analyze a classical universe
filled with exotic dust, which means $\rho_m<0$, i.e $p_{\varphi}<0$.
For simplicity, let us focus on the flat spatial case.

In the presence of exotic dust, the behaviour of the scale factor is drastically
different. From the Friedmann Eq.(\ref{friedmann}), since $p_{\varphi}<0$,
the radiation density must always be equal or greater then the dust density,
otherwise the Friedmann equation
\[
\left(\frac{\dot{a}}{a}\right)^{2}=\frac{1}{6}\left(\frac{p_{\eta}}{a^{2}}-
\frac{\mid p_{\varphi}\mid }{a}\right)
\]
has no solution.
For small values of the scale factor, the radiation term dominates.
As the scale factor grows, the exotic dust term begins to be comparable
to the radiation term up to the critical point where both are equal and $\dot{a}=0$.
>From this point, the scale factor decreases until the universe recollapses.
Note that Eq.(\ref{cdr}) for $\kappa=0$ and $p_{\varphi}<0$ implies that $a''<0$
at all times.
Hence, contrary to the normal dust matter case where after the big bang
the universe expands forever [see Eq.(\ref{cdr}) for $\kappa=0$], in the exotic case the
universe recollapses in a big crunch. The qualitative evolution of the scale
factor is plotted in figure 1:
\begin{figure}[h]
\rotatebox{-90}{\includegraphics[width=7cm]{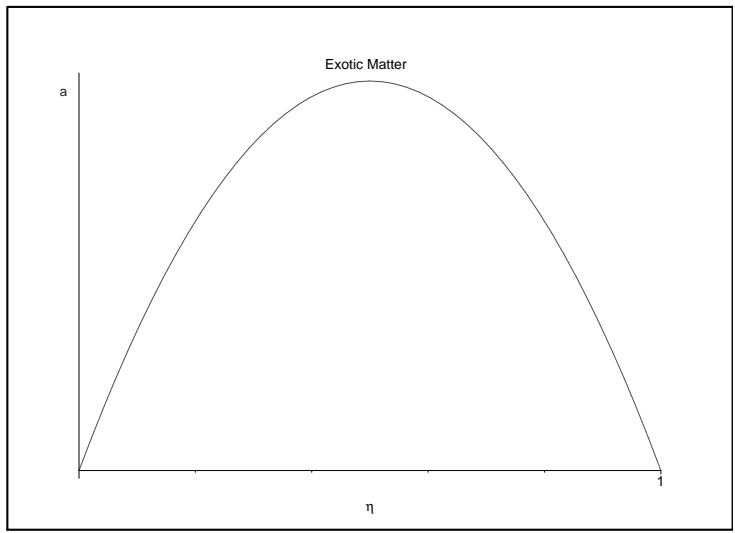}}
\caption{Scale factor evolution of a universe filled with exotic dust ($p_{\varphi}<0$).}
\end{figure}
The deceleration parameter in conformal time is given by
\begin{equation}\label{q}
q=-\frac{{a''}a}{{a'}^{2}}+1 .
\end{equation}
It diverges when the scale factor reaches its maximum value (${a'}=0$ and $a''<0$).

\section{FLRW Quantum Model with Radiation}\label{1f}

In this section, we present a general result concerning the presence of
singularities in the quantization of a FLRW model with radiation.
The hamiltonian constraint in this case is
\begin{equation}
\label{hconstr}
{\cal H}=-\frac{p_{a}^2}{24a}-6\kappa a+\frac{p_\eta}{a} \approx 0,
\end{equation}
and $\eta$ is conformal time, as discussed above.

Using the Dirac quantization procedure, the hamiltonian constraint
phase space function $\cal{H}$
becomes an operator which must annihilate the quantum wave function:
$\hat{\cal{H}}\Psi=0$. One then obtains in natural units
the Wheeler-De Witt equation for the minisuperspace FLRW
metric with radiation:
\begin{equation}
\label{sch27}
i \frac{\partial }{\partial \eta}\Psi\left(a,\eta\right)=
\left(-\frac{1}{24}
\frac{\partial^{2}}{\partial a^{2}}+6\kappa a^{2}\right)\Psi \left(a,\eta\right).
\end{equation}
Note that a particular factor ordering has been chosen and, because
$p_\eta$ appears linearly in Eq.(\ref{hconstr}), $\eta\rightarrow -\eta$ is chosen to be the
time label in which the wave function evolves (the sign reversing was
done in order to express this quantum equation in a familiar
Schr\"oedinger form \cite{Lapshinskii}).

The scale factor is defined only in the half line $[0,\infty)$, which means that
the superhamiltonian (\ref{hconstr}) is not in general hermitian. Hence, if one
requires
unitary evolution, the Hilbert space is restricted to functions in
$L^{2}(0,\infty)$ satisfying the condition
\begin{equation}\label{cond1}
\frac{\partial\Psi}{\partial a}(0,\eta)=\alpha\Psi(0,\eta),
\end{equation}
where $\alpha$ is a real parameter \cite{JMP371449}.

We will now show that condition (\ref{cond1}), together with the
assumption that $\Psi(a,\eta)$ is analytic in $\eta$ at $a=0$,
implies that general quantum solutions
of Eq.(\ref{sch27}), when interpreted using the BdB interpretation,
yield quantum cosmological models without any singularity.

We can rearrange Eq.(\ref{sch27}) in order to isolate the second spatial derivative:
\begin{equation}\label{d2psi}
\frac{\partial^{2}}{\partial a^{2}}\Psi \left(a,\eta\right)=
24\left[-i\frac{\partial }{\partial \eta}\Psi\left(a,\eta \right)+
6 \, \kappa \, a^{2}\Psi \left(a,\eta\right)\right].
\end{equation}
Using the BdB interpretation, the scale factor equation of motion is given by
the gradient of the phase $S\left(a,\eta\right)$ of the wave function
\[
a'=\frac{1}{12}S_a\left(a,\eta\right)=-\frac{i}{24}\frac{(\Psi \Psi_a^{ \ast }-\Psi_a
\Psi^{\ast })}{\Psi \Psi^{\ast }}\equiv f\left(a,\eta\right),
\]
where the index $a$ means derivative with respect to $a$.
Taking the boundary condition (\ref{cond1}) at $a=0$,
the velocity function $f\left(0,\eta\right)$ vanishes. Hence, if there is a time
$\eta_0$ where $a(\eta_0)=0$, then  $a'(\eta_0)=0$. For $a''$ one has:
\[
{a''}=\frac{\partial f}{\partial a}a'+
\frac{\partial f}{\partial \eta}=\frac{\partial f}{\partial a}f+
\frac{\partial f}{\partial \eta}.
\]
This is also zero at $a=0$ unless $\partial f/\partial a$ diverges there.
However,
\begin{eqnarray*}
 \frac{\partial f}{\partial a} & = & -\frac{i}{24}\frac{(\Psi \Psi_{aa}^{ \ast }-
 \Psi_{aa} \Psi^{\ast })}{\Psi \Psi^{\ast }}+\frac{i}{24}
 \frac{\biggl[\left(\Psi \Psi_a^{ \ast }\right)^{2}-
 \left(\Psi_a \Psi^{\ast }\right)^{2}\biggr]}
 {\left(\Psi \Psi^{\ast }\right)^{2}}\\
 & = &\frac{(\Psi
 \frac{\partial \Psi^{ \ast }}{\partial t}+\frac{\partial \Psi}{\partial t})
 \Psi^{\ast }}{\Psi \Psi^{\ast }}
-\frac{i}{2}\frac{\biggl[\left(\Psi \Psi_a^{ \ast }\right)^{2}-
\left(\Psi_a \Psi^{\ast }\right)^{2}\biggl]}{\left(\Psi \Psi^{\ast }\right)^{2}}\\
\end{eqnarray*}
is obviously finite if condition (\ref{cond1}) and analyticity of $\Psi$ in $\eta$
is satisfied at $a=0$.  The case when $\left(\Psi \Psi^{\ast }\right)^{2}=0$
does not need to be analyzed because bohmian trajectories cannot pass
through nodal regions of the wave function.

The same reasoning can be used for all higher derivatives $d^n a/d\eta^n$ at
$a=0$ to show that they are all zero: one just have to
use equation (\ref{d2psi}) to substitute $\partial ^2\Psi/\partial a^2$ for
$\partial \Psi/\partial \eta$ and condition (\ref{cond1}) to substitute
$\partial ^2\Psi/\partial a\partial \eta$
for $\alpha\partial \Psi/\partial \eta$ at $a=0$ whenever they appear,
and then use analyticity of $\Psi$ in $\eta$ at $a=0$.

With these results, if there is a time
$\eta_0$ where $a(\eta_0)=0$, expanding $a(\eta)$ in Taylor series around
$\eta_0$ shows that $a(\eta)\equiv 0$. This means that the only singular
bohmian trajectory is the trivial one of not having a universe at all!
All non trivial quantum solutions have to be non singular.

\section{Quantum behaviour of a FLRW Model With Dust and Radiation}
\label{quantum}

As we have seen in section \ref{drs}, the superhamiltonian constraint
for a FLRW model with
non interacting dust and radiation is given by Eq.(\ref{ham0}):

\begin{equation}\label{ham27}
{\cal{H}}\equiv -\frac{p_{a}^2}{24a}-6\kappa a + \frac{p_{\eta}}{a} + p_{\varphi}\approx 0,
\end{equation}
We see that both $p_\eta$ and $p_{\varphi}$ appear linearly in $\cal{H}$,
and their canonical coordinates $\eta$
and $\varphi$ are, respectively, conformal and cosmic time.
As in the preceeding section, from the Dirac quantization procedure
one obtains the quantum equation $\hat{\cal{H}}\Psi=0$,
which reads

\begin{equation}\label{hamo}
\left(\frac{1}{24a}\frac{\partial^{2}}{\partial a^2}-6\kappa a
-\frac{i}{a}\frac{\partial }{\partial \eta}-
i\frac{\partial }{\partial \varphi}\right)\Psi(a,\varphi , \eta)=0,
\end{equation}
where we have used the usual coordinate representation
$\hat{p}=-i\partial/\partial q$.

Either $\eta$ or $\varphi$ can be chosen as time parameters on which $\Psi$ evolves.
However, the classical solutions can be expressed explicitly only in
conformal time $\eta$ [see Eq.(\ref{cdr})]. Furthermore, cosmic time
$\varphi$ depends on the constants characterizing each particular solution
through $\varphi=\int d\eta a(\eta)$, and it is not the same parameter for all
classical solutions (see Ref.\cite{Tipler} for deatils). Hence,
we will take $\eta$ (in fact $-\eta$, for the reasons mentioned in
the previous section) as the time parameter of the quantum theory\footnote{The
choice of $\varphi$ will probably yield a different theory, with a different
Hilbert space. The kinetic term is more complicate and the measure is not
the trivial one. We will not study this possibility here.}.
With this choice, and for a particular factor ordering, Eq.(\ref{hamo}) can
be written as:
\begin{equation}\label{hamo2}
i\frac{\partial }{\partial \eta}\Psi(a,\varphi , \eta)=
\left(-\frac{1}{24}\frac{\partial^{2}}{\partial a^2}+6\kappa a^2
+i a\frac{\partial }{\partial \varphi}\right)\Psi(a,\varphi, \eta).
\end{equation}

\subsection{Eigenstates of total matter content}\label{the}

In this subsection we only consider initial states $|\Psi(\eta_0)\rangle$
which are eigenstates of the total dust matter operator $\hat{p}_{\varphi}$.
It follows that these states at  time  $\eta$, $|\Psi(\eta)\rangle$ will also be eigenstates
of $\hat{p}_{\varphi}$ with the same eigenvalue because $[\hat{H},\hat{p}_{\varphi}]=0$.
In other words, we consider that dust matter is not created nor destroyed.
In such a way, we have $\hat{p}_{\varphi}|\Psi(\eta)\rangle=p_{\varphi}|\Psi(\eta)\rangle$ and
the wave function in the  $a$, $\varphi$ representation,
$\langle a,\varphi|\Psi(\eta)\rangle=\Psi(a,\varphi,\eta)$, is given by

\begin{equation}\label{mom}
\Psi(a,\varphi,\eta)=\Psi(a,\eta)e^{ip_{\varphi} \varphi}.
\end{equation}
>From the Schr\"odinger's equation (\ref{hamo2}), we have for $\Psi(a,\eta)$
\begin{equation}\label{hamoeig}
i\frac{\partial }{\partial \eta}\Psi(a,\eta)=\left(-\frac{1}{24}
\frac{\partial^{2}}{\partial a^2}+ 6\kappa a^2-p_{\varphi}a\right)\Psi(a,\eta),
\end{equation}
which is the Schr\"odinger equation for a particle of mass $m=12$ in a one dimensional
forced oscilator with frequency $w=\sqrt{\kappa}$ and constant force $p_{\varphi}$,
which we write as

\begin{equation}\label{hamof}
i\frac{\partial}{\partial \eta}\Psi(a,\eta)=\left(-\frac{1}{2m}
\frac{\partial^{2}}{\partial a^2}+\frac{mw^2}{2}a^2-p_{\varphi}a\right)\Psi(a,\eta)
\end{equation}

The scale factor is defined only in the half line $[0,\infty)$, which means that the
hamiltonian (\ref{ham27}) is not in general hermitian. Hence, if one requires unitary
evolution, the Hilbert subspace is resctricted to functions on
$L^{2}(0,\infty;-\infty,\infty)$ satisfying the condition:
\begin{equation}\label{condition}
\int_{-\infty}^{\infty}{d\varphi \left[\frac{\partial \Psi^{\ast }_{2}
\left(a,\varphi , \eta\right)}{\partial a}\,
\Psi_{1}\left(a,\varphi , \eta\right)\right]_{a=0}}=\int_{-\infty}^{\infty}{d\varphi
\left[\frac{\partial
\Psi_{1}\left(a,\varphi , \eta\right)}{\partial a}\, \Psi^{\ast
}_{2}\left(a,\varphi , \eta\right)\right]_{a=0}}
\end{equation}
for any $\Psi_{1}(a,\varphi , \eta), \Psi_2(a,\varphi , \eta) \in L^{2}(0,\infty;-\infty,\infty)$.
In the special case considered in this section, this condition is reduced to

\begin{equation}\label{cond}
\frac{\partial\Psi}{\partial a}(0,\eta)=\alpha\Psi(0,\eta),
\end{equation}
where $\alpha$ is a real parameter \cite{JMP371449}. We will analyze the two
extreme cases: $\alpha=0$ and $\alpha=\infty$, which are the simpler and usually
studied in the literature on quantum cosmology
\cite{bola2,Lapshinskii,alvarenga,dewitt,gotay,JMP371449}.

For the case $\alpha=0$ we have that
\begin{equation}\label{alfa0}
\frac{\partial\Psi}{\partial a}(0,\eta)=0,
\end{equation}
which is satisfied for even functions of $a$. For the  case $\alpha=\infty$, we have
\begin{equation}
\Psi(0,\eta)=0   ,
\end{equation}
which is satisfied for odd functions of $a$.
Both of them address the boundary conditions of the wave packet near the singularity
at $a=0$.

In order to develop the BdB interpretation, we substitute into the Schr\"odinger's
equation (\ref{hamof}), the wave function in the polar form $\Psi=Ae^{iS}$,
obtaining for the real part

\begin{equation}\label{equacaoH-J1}
\frac{\partial S}{\partial t}+\frac{1}{2m}\left(\frac{\partial S}{\partial
a}\right)^{2}-a\, p_{\varphi}+
\frac{m\,w^{2}}{2}\,a^{2}+Q=0,
\end{equation}
where

\begin{equation}
Q\equiv -\frac{1}{2m\,{A}}\frac{\partial^{2}{A}}{\partial a^{2}}
\end{equation}
is the quantum potential.
The Bohm guidance equation reads
\begin{equation}\label{bgr}
ma'=\frac{\partial S}{\partial a}.
\end{equation}

A solution $\Psi(a,\eta)$ of Eq.(\ref{hamof}) can be obtained from an initial wave
function $\Psi_{0}(a)$ using the propagator of a forced harmonic oscillator.
Let us do it for the two boundary conditions just presented.

\subsubsection{\bf The case of boundary condition $\alpha=0$}\label{case1}

We denote the propagator  $K^{\alpha=0}(2,1)\equiv K^{\alpha=0}(\eta_2,a_2;\eta_1,a_1)$,
where
$1$ stands for the initial time and initial scale factor $\eta_1, a_1$ respectively,
and $2$ stands for their final values.

The propagator when the Hilbert space is restricted to $a>0$ can be obtained
from the usual one (i.e with coordinate $-\infty<a<\infty$)  which is associated to a
particle in  a forced oscilator $K(2,1)\equiv K(\eta_2,a_2;\eta_1,a_1)$ by means of
\begin{equation}\label{Kpar}
K^{\alpha=0}(2,1)=K(\eta_2,a_2;\eta_1,a_1)+K(\eta_2,a_2;\eta_1,-a_1)
\end{equation}
This symmetry condition is necessary to consistently eliminate the contribution of
the negative values of the scale factor \cite{IJMPA53029}.
The  usual propagator associated to a particle in  a forced oscilator is \cite{feynman}:

\begin{equation}
K(2,1)=\sqrt{\frac{mw}{2 \pi  i \sin(w\eta)}} \exp(i \, S_{cl})
\end{equation}
where $\eta\equiv \eta_2 - \eta_1$ . The classical action $S_{cl}$
is given by

\begin{eqnarray}
 S_{cl}&=&\frac{mw}{2\sin(w\eta)} \biggl\{\cos(w\eta)(a_{2}^2+a_{1}^2)-2a_{2}a_{1}+(a_{2}+
 a_{1})\frac{2p_{\varphi}}{m w^2}[1-\cos(w\eta)]- \nonumber \\
&&[1-\cos(w\eta)]\frac{2 p_{\varphi}^2}{m^2 w^4}+\frac{p_{\varphi}^2}{m^2 w^4}\sin(w\eta)
w\eta \biggr\} .
\end{eqnarray}
We assume that for $\eta_1=0,$ the  initial wave function  is given by
\begin{equation}\label{initwave}
\Psi_0(a)=\biggl(\frac{8\sigma}{\pi}\biggr)^{1/4}\exp(-\sigma a^2),
\end{equation}
where $\sigma>0$.
The wave function in a future time $\eta_2$ is

\begin{equation}
\Psi(a_2, \eta_2)=\int_{0}^{\infty} K^{\alpha=0}(2,1)\Psi_0(a_{1})da_{1}=
\int_{-\infty}^{\infty} K(2,1)\Psi_0(a_{1})da_{1},
\end{equation}
where the even caracter of $\Psi(a,0)$ has been taken into account to extend the integral.
Integrating and renaming $\eta\equiv\eta_2,\,a\equiv a_2$ we have

\begin{eqnarray}\label{psi1t}
\Psi^{\alpha=0}(a,\eta)=\biggl(\frac{8 \sigma}{\pi}\biggr)^{1/4}
\sqrt{\frac{mw}{i\cos(w\eta)
[2 \sigma   \tan(w\eta)-imw]}} \exp\biggr\{\frac{imw}{2  \tan(w\eta)}
\biggl[a^2 && + \nonumber \\
i\frac{mw}{\cos^2(w\eta)[2 \sigma   \tan(w\eta)-imw]}
\biggl(-a+\frac{p_{\varphi}}{m w^2}[1-\cos(w\eta)]\biggr)^2+\frac{2ap_{\varphi}}{m}
\frac{[1-\cos(w\eta)]}{w^2 \cos(w\eta)}+\nonumber \\
\frac{2 p_{\varphi}^2}{m^2} \biggl(\frac{[\cos(w\eta)-1]}{w^4 \cos(w\eta)} +
\eta\frac{\tan(w\eta)}{w^3}\biggr)\biggr]\biggr\}  .
\end{eqnarray}

\underline{Flat spatial section ($\kappa=0$)}\label{0}

We consider the case $\kappa=0$, which is obtained by taking the limit of the wave
function given by Eq. (\ref{psi1t}) for $w\rightarrow 0$. We compute its phase
$S$ from
$\Psi\equiv {\it A} e^{iS}$ and calculate the
derivative $\partial S/\partial a$. In this way we have, for the
Bohm guidance equation Eq.(\ref{bgr}),
\begin{equation}
a'-\frac{4\sigma^2 \eta}{4\sigma^2 \eta^2+m^2}a=
\frac{1}{m}\frac{(2\sigma^2 \eta^2+m^2)}{(4\sigma^2 \eta^2+m^2)}p_{\varphi} \eta .
\end{equation}
Comparing with the radiation case studied in \cite{bola2}, we see that here it appears a
term proportional to $p_{\varphi}$ in the RHS of the Bohm equation. The general
solution is:

\begin{equation}\label{gk0}
a(\eta)=C_0 \sqrt{4\sigma^2 \eta^2+m^2}+\frac{p_{\varphi}}{2m}\eta^2 ,
\end{equation} where $C_0$ is a positive integration constant. We can see that,
contrary to the classical solution (\ref{cdr}),
there is no singularity at $\eta=0$.
The quantum effects avoid it. Furthermore, for
long times $\eta\gg m/2\sigma$, Eq. (\ref{gk0}) reproduces the classical
behaviour (\ref{cdr}) for the scale factor.

For the case in which the evolution starts from a shifted gaussian wave function

\begin{equation}
\Psi_0(a)=\biggl(\frac{8\sigma}{\pi}\biggr)^{1/4}\exp\left[-\sigma (a-a_0)^2\right],
\end{equation}
the Bohm guidance relation contains an additional term yielding the
general solution
\begin{equation}
a(\eta)=C_0 \sqrt{4\sigma^2 \eta^2+m^2}+\frac{p_{\varphi}}{2m}\eta^2+\frac{a_0}{2},
\end{equation}
which has exactly the same behaviour, apart from the shift on the minimal value of the
scale factor by the $a_0/2$ term.

\underline{ Positive curvature spatial section ($\kappa =1$)}

Setting $w=\sqrt{\kappa}=1$ in the wave function given by Eq. (\ref{psi1t})
and computing its phase $S$, we obtain for the bohmian trajectories
\begin{equation}\label{gk1}
a(\eta)=C_0 \sqrt{4\sigma^2\sin ^2(\eta)+m^2\cos ^2(\eta)}+
\frac{p_{\varphi}}{2m}[1-\cos(\eta)],
\end{equation}
where  $C_0$ is a positive integration constant. This is a
non singular cyclic universe, see figure 2, which presents classical behaviour
for $\eta$ such that $\mid \tan (\eta)\mid\gg m/2$ [see Eq. (\ref{cdr})].
Quantum effects avoid the classical big bang and big crunch.

\begin{figure}[h]
\rotatebox{-90}{\includegraphics[width=7cm]{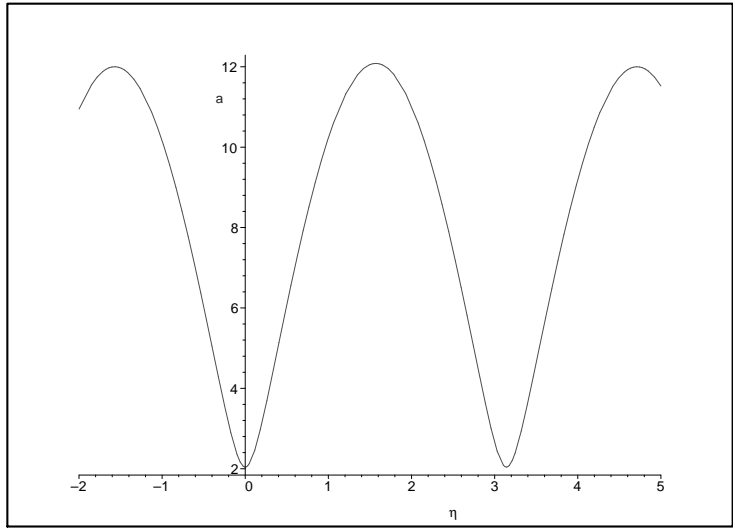}}
\caption{\small{Scale factor evolution of a quantum universe with positive curvature
spatial section filled with  matter and radiation.}}
\end{figure}

\underline{Negative curvature spatial section  ($\kappa=-1$)}

Setting now $w=\sqrt{\kappa}=i$ in the wave function (\ref{psi1t})
yields the bohmian trajectories:

\begin{equation}\label{gk-1}
a(\eta)=C_0 \sqrt{4\sigma^2\sinh ^2(\eta)+m^2\cosh ^2(\eta)}+
\frac{p_{\varphi}}{2m}[\cosh(\eta)-1].
\end{equation}
Again, $C_0$ is a positive integration constant. This is a
non singular ever expanding universe which presents classical behaviour
for $\eta$ such that $\mid \tanh (\eta)\mid\gg m/2$ [see Eq. (\ref{cdr})].
Quantum effects avoid the classical big bang.

As in the $\kappa =0$ case, a shift in the center of the initial gaussian
will not modify these solutions qualitatively.

For the boundary conditions $\alpha =\infty$, or $\Psi(0,t)=0$,
the propagator  $K^{\alpha=\infty}(2,1)$ can be obtained from the usual
(i.e, with coordinate $-\infty<a<\infty$)  propagator
associated to a particle in  a forced oscilator $K(2,1)$ by means of

\begin{equation}\label{Kimpar}
K^{\alpha=\infty}(2,1)=K(\eta_2,a_2;\eta_1,a_1)-K(\eta_2,a_2;\eta_1,-a_1) .
\end{equation}
In order to satisfiy the condition $\Psi(0,\eta)=0$, we take as the initial wave
function a wave packet given by

\begin{equation}
\Psi_0(a)=\biggl(\frac{128 \sigma^3}{\pi}\biggr)^{1/4}a \exp(-\sigma a^2) ,
\end{equation}
where $\sigma>0$. Following a similar procedure as in the case $\alpha=0$,
we calculate the wave function
by propagating the initial wave function as

\begin{equation}
\Psi(a_{2}, \eta_{2})=\int_{0}^{\infty} K^{\alpha=\infty}(2,1)\Psi_0(a_{1})da_{1}=
\int_{-\infty}^{\infty} K(2,1)\Psi_0(a_1)da_{1} ,
\end{equation}
where the odd caracter of $\Psi$ has been used in order to extend the integral.
Integrating this expression and renaming $a\equiv a_{2}$ and $\eta \equiv \eta_{2}$ with $\eta_1=0$, we have

\begin{equation}
\Psi^{\alpha=\infty}(a,\eta)=\biggl(\frac{-C}{2 D}\biggr)\Psi^{\alpha=0}(a,\eta)
\end{equation}
where

\begin{equation}
C\equiv \frac{imw}{\sin(w\eta)}\biggl[-a+\frac{p_{\varphi}}{mw^2}(1-\cos(w\eta)\biggr]
\end{equation}
and

\begin{equation}
D\equiv \frac{imw}{2\tan(w\eta)}-\sigma
\end{equation}
The phase of $\Psi^{\alpha=\infty}(a,\eta)$ can be expressed as the sum :

\begin{equation}
{\rm phase}[\Psi^{\alpha=\infty}(a,\eta)]=
{\rm phase}\biggl(\frac{-C}{2 D}\biggr) + {\rm phase}
[\Psi^{\alpha=0}(a,\eta)],
\end{equation}
and it is easy to see that the phase of $(-C/2 D)$ is independent of $a$. Then,
$[\partial {\rm phase}(\Psi^{\alpha=\infty}(a,\eta))]/\partial a=
[\partial{\rm phase} [\Psi^{\alpha=0}(a,\eta)]]/\partial a$, and the Bohm guidance
relations are the same as in the previous cases.
Therefore, the solutions are the same.

The quantum cosmological models obtained in this subsection have the nice properties
of being non singular
and presenting classical behaviour for large $a$. However, they suffer from a
fundamental problem: the wave function (\ref{psi1t}) from which they are obtained
does not have an unitary evolution. The reason is that propagators constructed from Eq.'s (\ref{Kpar}) and (\ref{Kimpar})
do not in general preserve the hermiticity condition (\ref{cond}) imposed on the wave
functions: it depends on the classical potential.
In Ref.\cite{IJMPA53029}, there are obtained the potentials which allow propagators
in the half line ($a>0$) to preserve unitary evolution. The potentials of the previous
section
are some of them but the potentials of the present
one are not. Hence, even though
the initial wave function Eq.(\ref{initwave}) satisfies the hermiticity condition, the
wave function (\ref{psi1t}) does not. Let us then explore the more general
case of initial superpositions of the total dust mass operator eigenstates.

\subsection{Analysis of wave packets given by superpositions of total dust
mass eigenstates}

In this subsection we consider the case of a general solution of Eq.(\ref{hamo2})
which is not necessarily  one of the  eigenstates of $\hat{p}_{\varphi}$, the total
dust mass operator.

Following the BdB interpretation of quantum mechanics, we substitute in
Eq.(\ref{hamo2}) the wave function in polar form: $\Psi = A\left(a,\varphi,\eta \right)
\exp\left\{{i}S\left(a,\varphi,\eta \right)\right\}$.
The dynamical equation splits in two real coupled equation for the two real
functions $S$ and $ A$ (recall that $w=\sqrt{\kappa}$ and $m=12$).
\begin{equation}\label{equacaoH-J}
\frac{\partial S}{\partial \eta}+\frac{1}{2m}\left(\frac{\partial S}{\partial
a}\right)^{2}-a\frac{\partial S}{\partial \varphi}+
\frac{m\, w^{2}}{2} a^{2}+ Q = 0 ,
\end{equation}

\begin{equation}\label{equacaocontinuidade}
\frac{\partial  A^{2}}{\partial \eta}+\frac{\partial}{\partial \varphi}\left(a\,
 A^{2}\right)+\frac{\partial}{\partial
a}\left( A^{2}\frac{1}{m}\frac{\partial S}{\partial a}\right)=0 ,
\end{equation}
where
\begin{equation}
Q\equiv -\frac{1}{2m\, A}\frac{\partial^{2}{A}}{\partial a^{2}} .
\end{equation}

Equation (\ref{equacaoH-J}) is the modified Hamilton-Jacobi equation
where $Q\left(a,\varphi,\eta\right)$ is the quantum potential which is responsible for all
the peculiar non classical  behaviours. When the quantum potential is zero, the equation
is exactly the classical Hamilton-Jacobi equation. The momenta are given by the
Bohm's guidance equations
\begin{eqnarray}
p_{a}&\equiv & \frac{\partial S\left(a,\varphi,\eta\right)}{\partial a},\\
p_{\varphi}&\equiv& \frac{\partial S\left(a,\varphi,\eta\right)}{\partial \varphi} .
\label{bphi}
\end{eqnarray}
Note also that
\begin{equation}
\label{rad}
p_\eta=\frac{\partial S\left(a,\varphi,\eta\right)}{\partial \eta}
\end{equation}
is the total `energy' of the system, which is interpreted, from
its classical meaning, as the total amount of radiation in the universe model.

In the causal interpretation, equation (\ref{equacaocontinuidade}) is a continuity
equation where ${A}^{2}$ is a probability density. The generalised
velocities can easily be identified as
\begin{eqnarray}
{a'} &\equiv & \frac{1}{m}\frac{\partial S\left(a,\varphi,\eta\right)}{\partial a} ,
\label{ba}\\
{\varphi'}&\equiv & a .\label{velocphi}
\end{eqnarray}
Consider now the classical limit ($Q=0$). Then the solution of the principal
Hamilton function ($S$) is just $S=W\left(a\right)-E\eta+p_{\varphi}\varphi$,
where $E$ and $p_{\varphi}$
are constants. Since $p_{\varphi}$ is proportional to the total amount of dust matter in
the universe, and $E$ to the total amount of radiation, there is no creation or
annihilation of dust matter nor radiation.
However, in the presence of a quantum potential,
this solution is no longer valid, opening the possibility of non conservation of matter
and radiation due to quantum effects.

\subsubsection{Formal Solutions}
We now turn to the problem of solving the Schr\"odinger's equation (\ref{hamo2}).

\underline{Flat spatial section}

For the case of flat spatial section  ($\kappa=0$), equation (\ref{hamo2}) simplify to

\begin{equation}
i\frac{\partial \Psi \left(a,\varphi ,\eta\right)}{\partial \eta}=
-\frac{1}{2m}\frac{\partial^{2}
\Psi \left(a,\varphi ,\eta\right)}{\partial
a^{2}} +i a\frac{\partial \Psi \left(a,\varphi ,\eta\right)}{\partial \varphi}
\label{eqkzero}
\end{equation}
To solve this equation we make the
ansatz
\begin{equation}
\Psi \left(a,\varphi ,\eta\right)=\chi\left(a\right) \exp \left(-\frac{i}{2m}\beta \,
\eta\right) \exp\left(\frac{i}{2m}\upsilon \, \varphi\right) ,
\end{equation}
where $\chi\left(a\right)$ must satisfy the differential
equation\footnote{As in the following we will make superpositions of
eigenfunctions of the total dust matter operator, we will use from now on the letter
$\upsilon$ in order to not confuse it with the beable
$p_{\varphi}=\partial S/\partial\varphi$. We did not make this distinction
before because they coincide for eigenfunctions
of the total dust matter operator.}

\begin{equation}
\frac{\partial^{2} \chi\left(a\right)}{\partial a^{2}}+\upsilon a
\chi\left(a\right)+\beta\chi\left(a\right)=0.
\end{equation}
This is essentially an Airy equation with solution given by

\begin{equation}
\chi\left(a\right)=\sqrt{a+\frac{\beta}{\upsilon}}\left\{A\,Z_{\frac{1}{3}}
\left[\frac{2\sqrt{\upsilon}}{3}\left(a+\frac{\beta}{\upsilon}
\right)^{\frac{3}{2}}\right]+ B\,Z_{-\frac{1}{3}} \left[\frac{2\sqrt{\upsilon}}{3}
\left(a+\frac{\beta}{\upsilon}\right)^{\frac{3}{2}}\right]\right\}
\end{equation}
The $Z_{\frac{1}{3}}$ function is the first kind Bessel function of degree
$\frac{1}{3}$, and the $A$ and $B$ can be any functions of $\upsilon$ and $\beta$.

The general solution is a superposition given by
\begin{eqnarray*}
\Psi \left(a,\varphi,\eta\right)= \int{d\beta\, d\upsilon
\exp\left\{-\frac{i}{2m}\beta\,\eta\right\}\exp\left\{\frac{i}{2m}\upsilon\,
\varphi\right\}\sqrt{a+\frac{\beta}{\upsilon}}} \times \\
\times \left\{ A\left(\beta,\upsilon\right) \, Z_{\frac{1}{3}} \left[
\frac{2\sqrt{\upsilon}}{3}
\left(a+\frac{\beta}{\upsilon}\right)^{\frac{3}{2}}\right] +B\left(\beta,\upsilon\right)
\,Z_{-\frac{1}{3}}\left[\frac{2\sqrt{\upsilon}}{3}
\left(a+\frac{\beta}{\upsilon}\right)^{\frac{3}{2}}\right]\right\}\\
\end{eqnarray*}

\underline{Positive curvature spatial section: Landau levels}

In the positive curvature case ($\kappa=1$), Eq.(\ref{hamo2}) reads

\begin{equation}\label{wheeler-dewittk1}
i\frac{\partial \Psi \left(a,\varphi ,\eta\right)}{\partial \eta}=
-\frac{1}{2m}\frac{\partial^{2}
\Psi \left(a,\varphi ,\eta\right)}{\partial
a^{2}} +\frac{m}{2}a^{2} \Psi \left(a,\varphi,\eta\right)+i a\frac{\partial \Psi
\left(a,\varphi ,\eta\right)}{\partial\varphi} .
\end{equation}
There is a canonical transformation which simplifies the problem.
Let us define new variables given by
\begin{eqnarray*}
\xi \equiv \sqrt{m}\, a - \frac{p_{\varphi} }{\sqrt{m}} &; &  \sigma \equiv
-\sqrt{m}\, \varphi + \frac{p_{a}
}{\sqrt{m w}} ,\\
p_{\xi} \equiv \frac{p_{a}}{\sqrt{m}} &; & p_{\sigma} \equiv
-\frac{p_{\varphi}}{\sqrt{m}} .
\end{eqnarray*}
Using these new variables, the hamiltonian decouples in two parts, one describing
a harmonic oscillator and the other a free particle:

\begin{equation}\label{schk1}
\hat{H}=\underbrace{\frac{1}{2}\left(\hat{p}_{\xi}^{2}+
\hat{\xi}^{2}\right)}_{\mbox{\it harmonic
oscillator}}-\underbrace{\frac{1}{2}\hat{p}_{\sigma}^{2}}_{\mbox{ \it free particle}} .
\end{equation}
Decomposing the wave function as

\begin{equation}
\Psi\left(\xi,\sigma,\eta \right)=
\chi\left(\xi\right)\exp\left\{-i\left(\epsilon\,\eta+\sqrt{2\,k}\,\sigma\right)\right\},
\end{equation}
we immediately recognize that $\chi\left(\xi\right)=
\exp\left\{-\frac{\xi^{2}}{2}\right\}h_{n}\left(\xi\right)$, where $h_{n}$
are the Hermite polinomial of degree $n$. Just as for the harmonic oscillator,
the index $\epsilon$ is constrained to take the values
\begin{equation}\label{landau}
\epsilon_{n}=k+\left(n+\frac{1}{2}\right)  ,
\end{equation}
where $k$ can take any real positive value while $n$ is a positive integer.
Eq. (\ref{landau}) determines  a set of {\it Landau levels} for the cosmological model
\cite{landaulevels}. The most general solution is a superposition given by

\begin{eqnarray}\label{solucaogeral}
\Psi \left( \xi ,\sigma ,\eta \right) &=& \sum_{n=0}^{\infty}{
\int{dk\,\chi_{n}\left(\xi\right)\left[D_{n}\left(k\right)\exp\left\{i\sigma\,
\sqrt{2\,k}\right\}+ \right. }}  \nonumber \\
&& \left. G_{n}\left(k\right)\exp\left\{-i\sigma\,\sqrt{2\,k}\right\}\right]\times
\exp\left\{-i\, \epsilon_{n}\, \eta\right\}.
\end{eqnarray}
The quantities $D_{n}\left(k\right)$ and $G_{n}\left(k\right)$ are arbritary
coefficients that can depend on the parameter $k$. Recall that we have performed a
canonical transformation that mix coordinates and momenta, and these are not the
proper variables to apply the causal interpretation. Instead, it is imperatif to
apply the inverse transformation to the coordinate basis before using
the guidance relations. This is a necessary
requirement to maintain the consistency of the causal interpretation of quantum
mechanics \cite{CQG141993}-\cite{PR89319B}.\\

For the negative curvature spatial section ($\kappa=-1$),
the general solutions are hypergeometric functions
whose asymptotic behaviours are rather complicated to study
in order to obtain reasonable boundary conditions. Hence,
we will not treat this case here. We proceed to the analysis of
an interesting particular solution.

\subsubsection{Transition from exotic dust to dust in the flat case}

The quantum states of the matter and radiation FLRW universe studied in section \ref{0}
are eigenstates of the total dust matter operator ${\hat{p}}_{\varphi}$. The
total wave function is given by
$\Psi(a,\varphi,\eta)= \Psi(a,\eta) \exp(i\varphi p_{\varphi})$
where $\Psi(a,\eta)$ is given by Eq.(\ref{psi1t}).
Taking the
limit $w \rightarrow 0$ in that equation, we obtain the wave function $\Psi(a,\eta)$ for
the case of flat spatial section, $\kappa=0$ which, after renaming the eigenvalues of
total mass by $\upsilon\equiv p_{\varphi}$, is given by

\begin{eqnarray}
\Psi_{\upsilon}\left(a,\eta\right)= \left(\frac{8\sigma m^{2}}{\pi\,\mu}
\right)^{\frac{1}{4}}\exp\left\{
-\frac{m^{2}\sigma}{\mu}\left(a-\frac{\upsilon\,\eta^{2}}{2m}\right)^{2}-
i\frac{\upsilon^{2}\eta^{3}}{6m}-i\frac{\theta}{2}+\right.&  \nonumber \\
\left.+i\frac{m}{2  \eta}\left[
\left(a+\frac{\upsilon\,\eta^{2}}{2m}\right)^{2}-\frac{m^{2}}{\mu}
\left(a-\frac{\upsilon\,\eta^{2}}{2m}\right)^{2}\right] \right\} & ,
\end{eqnarray}
where
\begin{eqnarray*}
&&\mu= 4\sigma^{2}\eta^{2}+m^{2} ,\\
&&\theta= \arctan \left(\frac{2\sigma \eta}{m}\right) .
\end{eqnarray*}

Now we consider  a more general situation than in section \ref{0}.
We suppose an initial state at $\eta=0$ which is given by a gaussian superposition
of eigenstates of total matter

\begin{equation}\label{superposicaoguassiana0}
\Psi \left(a,\varphi,0\right)=\int_{-\infty}^{\infty}{d \upsilon\,
\exp^{-\gamma\left(\upsilon-\upsilon_{0}\right)^{2}}\Psi_{\upsilon}\left(a,0\right)\,
\exp\{-i\,\varphi\,\upsilon\}} .
\end{equation}
Then, the state at time $\eta$ is given by
\begin{equation}\label{superposicaoguassiana}
\Psi \left(a,\varphi,\eta \right)=\int^{\infty}_{-\infty}{d \upsilon \,
\exp^{-\gamma\left(\upsilon-\upsilon_{0}\right)^{2}}\Psi_{\upsilon}\left(a,\eta \right)\,
\exp\{-i\,\varphi\,\upsilon\}} .
\end{equation}
In this way, we have a square-integrable wave function. We find

\begin{eqnarray*}
\Psi \left(a,\varphi,\eta\right)= \left(\frac{8\sigma \pi
m^{2}}{\mu\,\nu}\right)^{\frac{1}{4}}\exp\left\{\left(\frac{\Re\left(F\right)}{4\nu}-\frac{\sigma
m^{2}}{\mu}\right)a^{2}+\frac{\Re\left(G\right)}{4\nu}\,a\,\varphi
+\frac{\Re\left(J\right)}{4\nu}\varphi^{2}+\frac{\Re\left(L\right)}{4\nu}\,a+\right. & \\
\left. +\frac{\Re\left(M\right)}{4\nu}\varphi +\frac{\Re\left(P\right)}{4\nu}\; + i\left[ \,
\left(\frac{\Im\left(F\right)}{4\nu}+\frac{m}{2\mu   \eta}\left(\mu-m^{2}\right)
\right)a^{2}+\frac{\Im\left(G\right)}{4\nu}a\, \varphi +\right. \right. & \\
\left.\left.  +\frac{\Im\left(J\right)}{4\nu}\varphi^{2}+\frac{\Im\left(L\right)}{4\nu}a
+\frac{\Im\left(M\right)}{4\nu}\varphi+\frac{\Im\left(P\right)}{4\nu} \right]-
i\frac{\theta+\tau}{2} ,
\right\} &
\end{eqnarray*}
where we defined

\begin{eqnarray*}
&&\nu= \left(\gamma+\frac{\sigma \eta^{4}}{4\mu}\right)^{2}+\frac{\eta^{6}}{\left(24m \mu
\right)^{2}}\left(\mu +3m^{2}\right)^{2}\\
&&\tau= \arctan \left[\frac{\eta^{3}(\mu+3m^{2})}{24m(\gamma
\mu+\sigma \eta^4)}\right]\\
&&F= \left[\frac{m\sigma \eta^{2}}{\mu}+i\frac{\eta}{2\mu  }\left(\mu+m^{2}\right)
\right]^{2}\left[\gamma+\frac{\sigma \eta^{4}}{4\mu}-i\frac{\eta^{3}}{24m\mu
 }\left(\mu+3m^{2}\right)\right]\\
&&G= -2\, i\left[\frac{m\sigma \eta^{2}}{\mu}+i\frac{\eta}{2\mu  }
\left(\mu+m^{2}\right)
\right]\left[\gamma+\frac{\sigma \eta^{4}}{4\mu}-i\frac{\eta^{3}}{24m\mu
 }\left(\mu+3m^{2}\right)\right]\\
&&J= -\left[\gamma+\frac{\sigma \eta^{4}}{4\mu}-i\frac{\eta^{3}}{24m\mu
 }\left(\mu+3m^{2}\right)\right]\\
&&L= -2\,i\,\gamma\upsilon_{0}\,G \\
&&M= 4\,i\,\gamma\upsilon_{0}  \,J \\
&&P= -4\gamma^{2}\upsilon_{0}^{2} \,J
\end{eqnarray*}
and $\Re$ and  $\Im$ stands for the real and imaginary part, respectively.

If one calculates the squared norm of the wave function, one obtains
\begin{equation}
\int_{0}^{\infty}{da}\int_{-\infty}^{\infty}{d\varphi}\left\|\Psi\right\|^{2}=
\sqrt{\frac{8\pi^{3}}{\gamma}}\left[1+\frac{1}{\sqrt{\pi}} {\rm{erf}}
\left(\frac{\upsilon_{0}
\eta^{2}}{2m}\right)\right],
\end{equation}
where ${\rm{erf}}(x)$ is the error function.
The only dependence on time can be eliminated by
choosing the gaussian to be centered at $\upsilon_{0}=0$. With this choice
we garantee unitary evolution of the total wave function.
>From equations
(\ref{bphi})-(\ref{velocphi}), the trajectories can be
computed by solving the given system of equations

\begin{eqnarray}
& &{a'}= \frac{2}{m} \left[\frac{\Im\left(F\right)}{4\nu}+\frac{m}{2\mu
\eta}\left(\mu-m^{2}\right)\right]\,a\,+ \frac{\Im\left(G\right)}{4m\nu}\,
\varphi \label{aevol}\\
& &{\varphi'}= a \\
& &p_{\varphi} = \left[ 2\frac{\Im\left(J\right)}{4\nu}\, \varphi
+\frac{\Im\left(G\right)}{4\nu}\,a\right]
\end{eqnarray}
Note that $p_{\varphi}$ is no longer constant.
We integrated numerically these equations with the
renormalisation condition $a\left(0\right)=1$.
The quantum potential $Q\equiv
-\frac{1}{2m\, A}\frac{\partial^{2} A}{\partial a^{2}}$
is non zero only close to the origin as shown on figure 3. Hence, we expect that quantum effects be relevant
only in this region. Far form the origin, the scale factor must behave classically.

\begin{figure}[!h]
\rotatebox{-90}{\includegraphics[width=7cm]{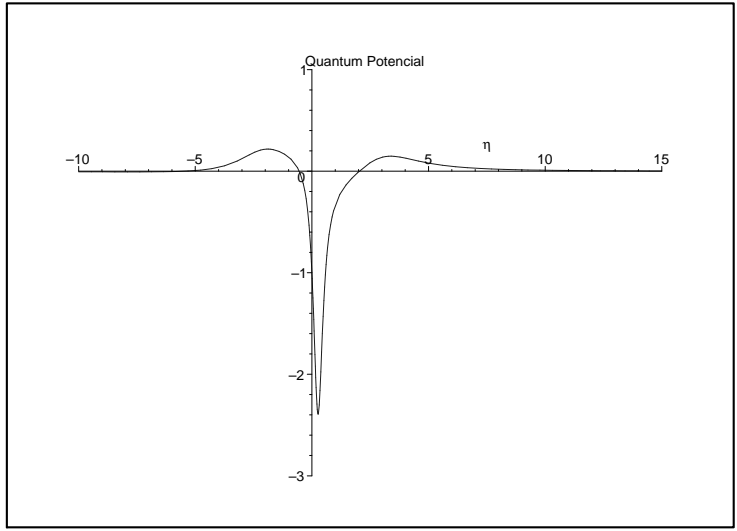}}
\caption{\small Quantum potential as a function of time. For regions far from
the origin, the quantum potential goes to zero, implying a classical behaviour.}
\end{figure}

The behaviour of $p_{\varphi}$ is plotted on figure 4.
\begin{figure}[!h]
\rotatebox{-90}{\includegraphics[width=7cm]{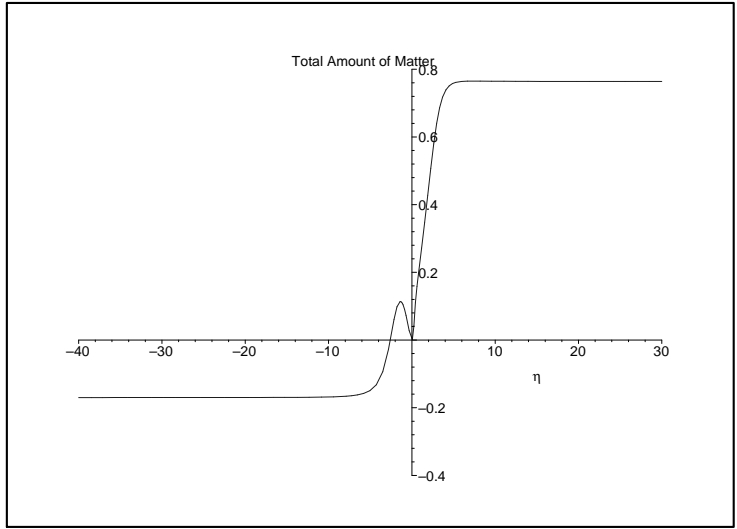}}
\caption{\small Evolution of the total amount of matter in the universe ($p_{\varphi}$).
In the far past, the universe was filled with exotic dust ($p_{\varphi}<0$). Close to the
origin, quantum effects transform it into conventional dust ($p_{\varphi}>0$).}
\end{figure}
>From this plot we can see that far from the origin $p_{\varphi}$ is constant.
This is in accordance with classical behaviour as long as the quantum potential is
zero in this region.
The surprising feature is that in the far past the universe was filled with
a classical exotic dust ($p_{\varphi} < 0$).

>From Eq.(\ref{rad}), one can also compute the amount of radiation.
Figure 5 shows the result. Again, far from the origin, radiation
is conserved while in the origin, due to quantum effects, it is
not conserved.

\begin{figure}[!h]
\rotatebox{-90}{\includegraphics[width=7cm]{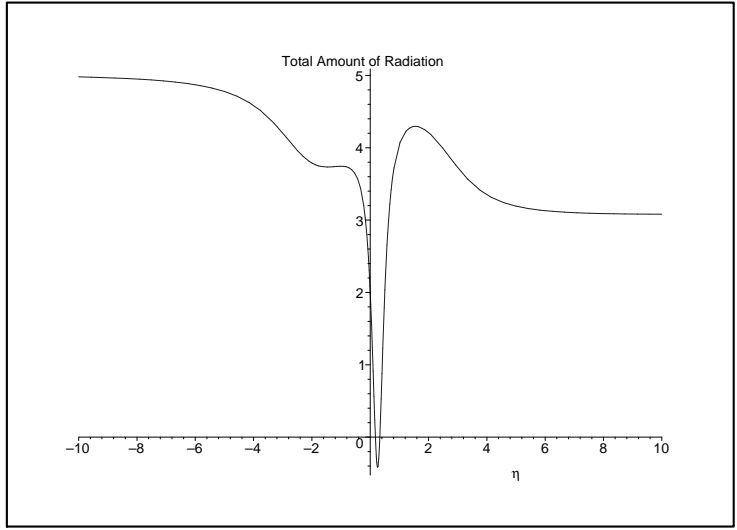}}
\caption{\small Evolution of the total amount of radiation in the universe ($p_{\eta}$).
Far from the origin the universe behaves classicaly and the total amount of
radiation is conserved. It varies near the origin due to quantum effects.}
\end{figure}

For the evolution of the scale factor,
numerical integration of equation (\ref{aevol}) yields the plot
of figure 6.
\begin{figure}[!h]
\rotatebox{-90}{\includegraphics[width=7cm]{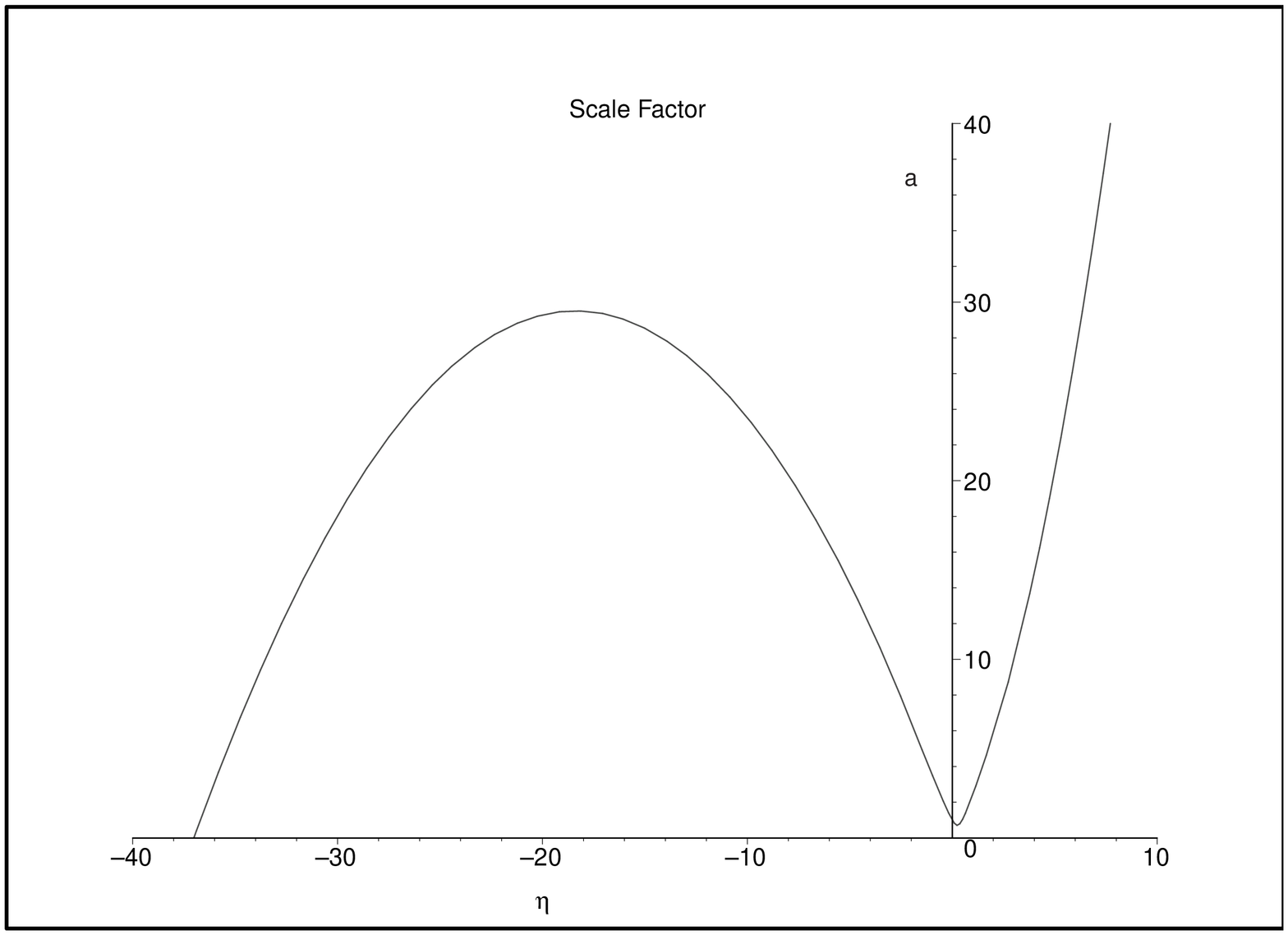}}
\caption{\small Evolution of the scale factor. Close the origin quantum effects
avoids the collapse. For the others regions, the evolution is essentially classical.}
\end{figure}
In the far positive region, the scale factor behaves classically, as expected, and matter
is conserved. On the other hand, in the far negative region the scale factor also behaves
classically but with a universe filled with exotic dust, and here again matter is
conserved (compare this region with figure 1). Both regions have a consistent
classical behaviour.
Hence, the universe begins classically from a big bang filled with exotic dust
and conventional radiation.
It evolves until it reaches a configuration when quantum effects avoid the classical
big crunch while transforming exotic dust into normal dust. From this point on,
the universe expands classically filled with conventional dust and radiation.

\section{Conclusions}\label{conclu}

In the present work we studied some features of the minisuperspace quantization
of FLRW universes with one and two fluids. For the one fluid case
(radiation), we have generalized results in the literature by showing that
all bohmian trajectories coming from reasonable general solutions of the
wave equation obtained through the assumptions of unitarity and analyticity at
the origin, do not present any singularity. Hence, this quantum minisuperspace
theory is free of singularities.

For the two fluids case (non interacting radiation and dust), we first obtained
bohmian quantum universes free of singularities reaching the classical
limit for large scale factors. However, these trajectories arise from eigenfunctions
of the total dust mass operator whose time evolution is not unitary. When considering
the general case, we managed to obtain a wave solution presenting unitary evolution
with some surprising
effects. Now dust and radiation can be created but the new feature is the possibility
of creation of exotic fluids. We have shown
that dust matter can be created as a quantum effect in such a way that the universe can
undergo a transition from an exotic dust matter era to a conventional dust matter one.
In this transition, one can see from figure 5 that radiation also becomes
exotic due to quantum effects, helping the formation of the bounce.

The fluid approach is not fundamental, but we expect that it can be quite
accurate in describing quantum aspects of the Universe, in the same way the
Landau description of superfluids in terms of fluid quantization was capable
of showing many quantum features of this system \cite{landau2}. After all,
creation and annihilation of particles as well as quantum states with negative energy
are usual in quantum field theory. The formalism developed in the present paper
seems to be a simple and calculable way to grasp these features of quantum field
theory. Their physical applications may be important: exotic fluids are relevant not
only in causing cosmological bounces and avoiding cosmological singularities
\cite{peter},
but also for the formation of wormholes \cite{thorn,matt} and for superluminal
travels \cite{27}. These are some developments of the present paper we
want to explore in future works.

\section*{ACKNOWLEDGEMENTS}

We would like to thank
{\it Conselho Nacional de Desenvolvimento Cient\'{\i}fico e Tecnol\'ogico}
(CNPq) of Brazil
and {\it Centro Latinoamericano de F\'{\i}sica} (CLAF) for financial
support.
We would also like to thank `Pequeno Seminario' of CBPF's Cosmology Group
for useful discussions.


\vspace{1.0cm}


\begin{thebibliography}{99}

\bibitem{bohm1} D. Bohm,  Phys. Rev. {\bf 85} (1952) 166.
\bibitem{bohm2} D. Bohm,  Phys. Rev. {\bf 85} (1952) 180.
\bibitem{hol} P. R. Holland, The Quantum Theory of Motion: An Account of the de Broglie-Bohm Causal Interpretation of Quantum Mechanics, (Cambridge University Press, Cambridge, 1993).
\bibitem{vink} J. C. Vink, Nucl. Phys. {\bf B369} (1992) 707.
\bibitem{bola1} J. A. de Barros and N. Pinto-Neto, Int. J. of Mod. Phys. {\bf D7} (1998) 201.
\bibitem{kow} J. Kowalski-Glikman and J. C. Vink, Class. Quantum Grav. {\bf 7} (1990) 901.
\bibitem{hor} E. J. Squires, Phys. Lett. {\bf A162}, (1992) 35.
\bibitem{bola2} J. A. de Barros,  N. Pinto-Neto and M. A. Sagioro-Leal, Phys. Lett.
{\bf A241} (1998) 229.
\bibitem{fab} R. Colistete Jr., J. C. Fabris and N. Pinto-Neto, Phys. Rev. {\bf D57} (1998) 4707.
\bibitem{fab2} R. Colistete Jr., J. C. Fabris and N. Pinto-Neto, Phys. Rev. {\bf D62} (2000) 83507.
\bibitem{must} N. Pinto-Neto and E. Sergio Santini, Phys. Rev.  {\bf D59}  (1999) 123517.
\bibitem{cons} N. Pinto-Neto and E. Sergio Santini, Gen. Relativ. Gravit.  34 (2002) 505.
\bibitem{tese} E. Sergio Santini, PhD Thesis, CBPF, Rio de Janeiro, May 2000,
gr-qc/0005092.
\bibitem{schutz1}B.F. Schutz, Phys. Rev.  {\bf D2} (1970) 2762 ; {\bf 4} (1971) 3559.
\bibitem{Misner} C. W. Misner, in: Magic Without Magic: John Archibald Wheeler, ed. J.R. Klauder, (Freeman, San Fransisco, CA, 1972).
\bibitem{Lapshinskii}V. G. Lapshinskii and V. A. Rubakov, Theor. Math. Phys. {\bf 33} (1977) 1076.
\bibitem{Tipler}F. J. Tipler, Phys. Rep. {\bf 137} (1986) 231.
\bibitem{alvarenga} F. G. Alvarenga, J. C. Fabris, N. A. Lemos and G. A. Monerat,
gr-qc/0106051.
\bibitem{dewitt} Bryce S. DeWitt, Phys. Rev.{\bf D} {\bf 160} 5 (1967)  1113.
\bibitem{feynman} R. P. Feynman and A. R. Hibbs, Quantum Mechanics and Path Integrals,
New  York , 1965.
\bibitem{gotay} M. J. Gotay and J. Demaret, Phys. Rev. {\bf D28} (1983) 2402.
\bibitem{landaulevels} L. D. Landau, Z. Phys. {\bf 64} (1930) 629, Reprinted and
translated in {\it Collected papers of Landau}, Paper 4, Edited by D. Ter Haar,
(Pergamon Press Ltd and Gordon and Breach, Science Publishers, Oxford, 1965).
\bibitem{incluir referencia} M.Abramowitz, I.A. Stegun-
``Handbook of Mathematical Functions'', (National Bureau of Standards,
Washington D.C., 1964).
\bibitem{VIIIBSCG299} N. Pinto-Neto, in: Cosmology and Gravitation II,
Proceedings of the VIII Brazilian School of Cosmology and Gravitation,
Edited by M\'ario Novello, (Editions Frontieres 1995).
\bibitem{JMP371449} N.A. Lemos, J. Math. Phys. {\bf 37} (1996) 1449.
\bibitem{IJMPA53029}E. Farhi, Int. J. Mod. Phy. A {\bf 5}  (1990) 3029.
\bibitem{CQG141993} J. Acacio de Barros, N. Pinto-Neto; Class. Quantum Grav. {\bf 14}
(1997) 1993.
\bibitem{PR89319B} D. Bohm; Phys. Rev. {\bf 89} (1952) 319.
\bibitem{landau2} I. M. Khalatnikov;  An Introduction to the Theory of
Super-fluidity (W. A. Benjamin, New York, 1965).
\bibitem{peter} P. Peter and N. Pinto-Neto, Phys. Rev. {\bf D65} (2001) 023513.
\bibitem{thorn} M. S. Morris and K. S. Thorne, Am. J. Phys. {\bf 56} (1988) 395.
\bibitem{matt} D. Hochberg, M. Visser, Phys. Rev. Lett. {\bf 81} (1998) 746.
\bibitem{27} K. D. Olum, Phys. Rev. Lett {\bf 81} (1998) 3567.

\end{thebibliography}
\end{document}